\documentclass[preprint,aps, nofootinbib]{revtex4}
\input{epsf.tex}
\usepackage[dvips]{graphics,epsfig}
\def\be{\begin{equation}}
\def\ee{\end{equation}}
\def\ba{\begin{eqnarray}}
\def\ea{\end{eqnarray}}

\begin{document}

\def\Box{\mathord{\dalemb{7.9}{8}\hbox{\hskip1pt}}}
\def\dalemb#1#2{{\vbox{\hrule height.#2pt
        \hbox{\vrule width.#2pt height#1pt \kern#1pt \vrule width.#2pt}
        \hrule height.#2pt}}}

\def\ba{\begin{eqnarray}}
\def\ea{\end{eqnarray}}
\def\be{\begin{equation}}
\def\ee{\end{equation}}
\def\tr{{\rm tr}}
\def\gtorder{\mathrel{\raise.3ex\hbox{$>$}\mkern-14mu
             \lower0.6ex\hbox{$\sim$}}}
\def\ltorder{\mathrel{\raise.3ex\hbox{$<$}\mkern-14mu
             \lower0.6ex\hbox{$\sim$}}}

\rightline{ORSAY LPT 04-118, DAMTP-2004-126}

\title{
Linearized Israel Matching Conditions for Cosmological
Perturbations in a Moving Brane Background
}

\author{
Martin Bucher${}^{1,3}$ 
and Carla Carvalho${}^{2,3}$ \\
${}^1$ Laboratoire de Physique Th\'eorique, 
Universit\'e Paris-Sud, 91405 Orsay, France\\
${}^2$ Departament de Fis\'\i ca Fondamental, Univsersitat de Barcelona,
08028 Barcelona, Spain \\
${}^3$ DAMTP, University of Cambridge, Cambridge CB3 0WA, United Kingdom
}

\begin{abstract}%
{
In the Randall-Sundrum cosmological models, a (3+1)-dimensional brane
subject to a $Z_2$ orbifold symmetry is embedded in a
(4+1)-dimensional bulk spacetime empty except for a negative
cosmological constant. The unperturbed braneworld cosmological solutions,
subject to homogeneity and isotropy in the three transverse spatial dimensions,
are most simply presented by means of a moving brane description. Owing to
a generalization of Birkhoff's theorem, as long as there are no
perturbations violating the three-dimensional spatial homogeneity
and isotropy, the bulk spacetime remains stationary and trivial. For
the spatially flat case, the bulk spacetime is described by one of
three bulk solutions: a pure $AdS^5$ solution, 
an $AdS^5$-Schwarzschild black hole solution,
or an $AdS^5$-Schwarzschild naked singularity solution. 
The brane moves on the
boundary of one of these simple bulk spacetimes, its trajectory
determined by the evolution of the stress-energy localized on it. 
We derive here the form of the Israel matching conditions
for the linearized cosmological perturbations in this moving brane picture. 
These Israel matching conditions must be satisfied in any gauge. However, they
are not sufficient to determine how 
to describe in a specific gauge the reflection of the bulk
gravitational waves 
off the brane boundary. In this paper we adopt a fully covariant Lorentz gauge
condition in the bulk and find the supplementary gauge conditions that
must be imposed on the boundary to ensure 
that the reflected waves do not violate the
Lorentz gauge condition. Compared to
the form obtained from Gaussian normal coordinates, the form of the
Israel matching conditions obtained here is more complex. However, the
propagation of the bulk gravitons is simpler because the
coordinates used for the background exploit fully the symmetry of the
bulk background solution.
}
\end{abstract}

\maketitle

\section{Introduction}

This paper discusses gravitational gauge invariance and gauge
fixing for the complete definition of the problem of computing the 
cosmological perturbations in a one-brane Randall-Sundrum scenario
\cite{rs}, \cite{gt1}.
In this cosmological scenario, the (4+1)-dimensional bulk spacetime
consists of a semi-infinite region, 
empty except for a negative cosmological constant, bounded by a
(3+1)-dimensional timelike surface on which a 
singular $Z_2$-symmetric distribution of stress-energy is localized.
The timelike boundary moves in the semi-infinite bulk spacetime,
its trajectory determined by the evolution of the stress-energy on it.
We assume that the bulk spacetime metric is obtained from a small 
linearized deviation 
from an exact solution to the Einstein equations,
so that $g_{ab} =g^{(0)}_{ab} +h_{ab}$ in the bulk.

In this paper we establish the form of the boundary
conditions coupling the degrees of freedom on the brane 
to those in the bulk when Lorentz gauge is chosen in the 
bulk. For the simplest situations such as a Minkowski
or de Sitter brane, this is not the simplest gauge 
choice. For these special situations it is advantageous
to employ Gaussian normal coordinates, so that $h_{55}$
and $h_{5\mu }$ vanish. However, for considering realistic
cosmological backgrounds, where the universe on the brane
has an essentially arbitrary expansion history, Gaussian 
normal coordinates are particularly awkward, and in 
many cases develop artificial coordinate singularities.
For studying more general, realistic cosmological solutions, it 
is advantageous to employ instead a ``moving brane"
description, where the unperturbed bulk remains static 
and trivial, described by coordinates such as the 
Poincar\'e coordinates
\be 
ds^2={1\over z^2}\left[ 
dz^2-dt^2+dx_1^2+dx_2^2+dx_3^2
\right] 
\ee
and the brane ``moves," tracing out a timelike trajectory
described as $z=z_b(t).$
The moving brane description has been developed in \cite{charmousis},
in contrast to the fixed brane approach used in \cite{binetruy}.
To consider perturbations in this moving boundary
description, it is advantageous to adopt a ``covariant"
gauge in the bulk, exploiting the full $AdS^5$ symmetry
of the unperturbed bulk. 
One such gauge is Lorentz gauge, where 
\be
\bar h_{ab;b}=\nabla _b \left[h_{ab} -{1\over 2}g^{(0)}_{ab}
\Bigl( g^{(0)~cd}~h_{cd}\Bigr) \right]=0.
\label{lgc-z}
\ee
In this gauge $h_{55}$ and $h_{\mu 5}$ no longer necessarily
vanish and the form of the linearized matching condition
contains additional terms. Moreover, five additional
auxilliary (``gauge") boundary conditions must be 
imposed at the brane. One might say that these are the 
boundary conditions for the five longitudinal (``pure gauge") 
modes present in $(4+1)$-dimensional Lorentz gauge. The
auxilliary gauge boundary conditions must be chosen in such 
a way that no forbidden polarization components are 
emitted or reflected into the bulk from the brane boundary. 

In this paper we establish the form of the linearized 
Israel matching condition and auxilliary gauge conditions
for Lorentz gauge.
In a forthcoming paper \cite{fcp} we apply the formal
results developed here to the actual problem
of computing the evolution of the cosmological perturbations
of the coupled brane-bulk system for various expansion
histories and various choices for the physical
degrees of freedom on the brane.

Linearized braneworld cosmological perturbations have been
previously studied by a large number of authors. 
(See refs.~\cite{l_start}--\cite{l_end} 
for a partial 
sampling of the literature.)
The majority of this work relies on the {\it pure 
scalar-vector-tensor} decomposition in the three
transverse spatial dimensions,
exploiting the three-dimensional homogeneity and
isotropy of the background solution, 
so that the problem problem may be decomposed into
separate sectors
that are either pure scalar, pure vector, 
or pure tensor 
with respect to the three transverse spatial dimensions.
As explained in \cite{l_start} and \cite{kodama}, generalizing
on previous work in \cite{gerlach}, each such sector contains 
only a single harmonic 
with respect to the three transverse spatial dimensions and 
its dynamics are described by a (1+1)-dimensional partial
differential equation in the $yt$-plane.
A sector corresponding to a pure tensor harmonic 
is described by a single coefficient function defined 
on the $yt$-plane, of scalar character with respect to
$y$ and $t,$ whose evolution is described by a $(1+1)$-dimensional 
partial differential equation. A sector corresponding to a pure 
vector harmonic has coefficient functions 
of both scalar and vector character with respect to $y$ and $t;$ 
however, as Mukohyama \cite{l_start} 
and Kodama et al.~\cite{kodama} have shown, these
coefficient functions can be expressed 
in terms of a single ``master" scalar function
of pure scalar character with respect to $y$ and $t$.
Similarly, for the sector corresponding to a 
pure scalar harmonic, the resulting coefficient functions 
are of scalar, vector, and tensor character
with respect to $y$ and $t.$ However, these in turn may be 
re-expressed in terms of another single ``master variable"
of scalar character.
The approach adopted in this paper differs in that the gauge
is fixed in a way that is completely local and 
independent of any particular choice of background coordinates 
or foliation of the background spacetime. It should be noted 
that the 
pure scalar-vector-tensor 
decomposition presupposes a high degree of symmetry and is
nonlocal.

The organization of the paper is as follows. In section 
\ref{Some Flat Space Examples} 
we analyse some simplified examples that can be solved
by expanding into plane waves in the bulk and individually
matching at the boundary. In particular, we 
consider reflection of electromagnetic waves
off a plane perfectly conducting wall under Lorentz gauge,
and also the reflection of linearized gravity waves off
a planar ``gravitational mirror" embedded in a flat
(i.e., Minkowski) background. A ``gravitational mirror"
is a $Z_2$-symmetric brane having a vanishing stress-energy
at zeroth order. 
Section \ref{Lorentz Gauge in a Curved Background Spacetime} 
derives the equation of motion of the bulk metric perturbations
in a general background in Lorentz gauge and characterizes
the residual gauge freedom and its relation to 
reparameterization invariance on the brane.
Section \ref{Compatibility of Sources on the 
Boundary with the Gauge Condition}  
considers the generalization of the results
of section \ref{Some Flat Space Examples} 
to branes curved at zeroth order embedded in 
bulk backgrounds curved at zeroth order. 
Section \ref{Stress-energy Conservation on the Brane at First Order} 
discusses stress-energy conservation on the brane and derives its
expression at linear order. 
In section \ref{The Linearized Israel Matching Condition and Its
  Divergence} 
the general form of the perturbative
Israel matching condition and its divergence is 
presented. This result is then applied to show that
the auxilliary boundary conditions proposed in 
section \ref{Compatibility of Sources on the 
Boundary with the Gauge Condition} 
are in fact consistent with Lorentz gauge. 
Finally, section \ref{Discussion} 
summarizes the results in a concise and self-contained form.
Details of the derivation of the evolution equations
of the bulk metric perturbations and the consistency
of Lorentz gauge in the bulk, of the linearized 
Israel matching condition, and of the extrinsic
curvature perturbations have been, respectively,
relegated to three appendices. 

Finally, we make explicit the notational conventions used in
this paper.
We use the metric signature $(-,+,+,+).$ 
Greek indices $\mu, \nu, \ldots $ denote 4-vectors in (3+1)
dimensions; lowercase latin indices denote 5-vectors in
a (4+1)-dimensional bulk spacetime (and in some cases, obvious
from the context, one of arbitrary
dimension as well). The uppercase latin indices
$A,B, \ldots $ denote directions parallel to a brane on the 
boundary of the bulk spacetime, and $N$ denotes the inward unit normal 
with respect to the boundary. 
The order of the 
derivatives with the semicolon convention is as
in the example, $W_{A;BC}=A^aC^cB^b\nabla _c\nabla _bW_a.$
The semicolon $;$ shall denote the covariant derivative with respect
to the bulk and $\vert $ denotes the covariant derivative
with respect to the brane.
We adopt the convention ${W^a}_{;bc}-{W^a}_{;cb}={R^a}_{dcb}W^d$
for the Riemann tensor.
On the boundary, we shall frequently use
Fermi normal coordinates with respect to the directions
on the surface of the brane, continuing these coordinates into the bulk 
using the Gaussian normal prescription. In this way
the connection coefficients vanish at a given point,
but in general not their partial derivatives. 
We shall also define $\nabla _A=(A^a\nabla _a)$ 
for the bulk covariant derivative and 
define $\tilde \nabla _A$ to be the corresponding derivation 
where the tilde indicates that the connection
is with respect to the induced metric on the brane.
We shall also sometimes suppress raising and lowering
of indices, it being implied that this is 
accomplished using the appropriate zeroth order metric.
Accordingly the Einstein convention of summing over repeated indices
is implied even when the pairs of repeated indices are 
both upper or both lower for more
orderly looking expressions.
We set $\kappa ^2=(8\pi G)$ where $G$ is Newton's gravitational
constant.

\section{Some Flat Space Examples: An Accelerating Conductor and Plane 
Gravitational Mirror in Minkowski Space}
\label{Some Flat Space Examples}

In this section we consider some very simple illustrations of
what is to follow in which electromagnetic and linearized
gravitational waves reflect from a flat planar boundary
embedded in flat Minkowski space.
In these simple cases the appropriate boundary conditions can
be established merely by considering a decomposition into
plane waves in the bulk. Later in the paper, when curved
boundaries and a curved bulk are considered, the technique
of expanding into plane waves can no longer be exploited.
Rather it is necessary to modify slightly the boundary
conditions obtained in this section to account for the 
curvature and to employ less direct arguments to demonstrate
rigorously their consistency with the bulk gauge condition.

\subsection{Reflection of electromagnetic waves off a planar
perfectly conducting boundary in Lorentz gauge}

\begin{figure}[t]
\setlength{\unitlength}{1cm}
\begin{center}
\begin{minipage}[t]{7.cm}
\begin{picture}(7.,7.)
\centerline 
{\hbox{\psfig{file=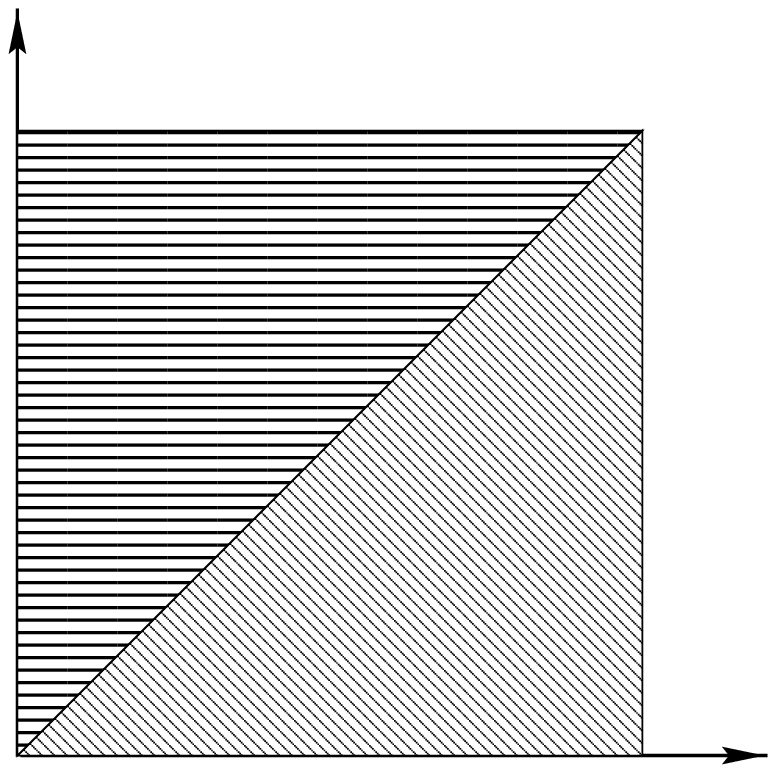,height=7.cm}}}
\put(-4,-0.5){$S $}
\end{picture}\par
\end{minipage}
\end{center}
\vspace{0.5cm}
\caption{\small \baselineskip=4pt {
{\bf Propagation of the residual gauge ambiguity.} 
Specifying initial data on the Cauchy surface $S ,$
taken here to coincide with $t=0,$ $z>0,$ completely specifies
the gauge in the diagonally hashed region. However, in the
horizontally hashed region, additional gauge information is
required to specify how the longitudinal pure gauge modes
scatter or propagate in from the timelike boundary at $z=0.$
}}
\label{Fig:ads}
\end{figure}

We consider classical electrodynamics
in Lorentz gauge in a
semi-infinite flat spacetime bounded by a planar accelerating 
perfectly conducting wall. 
For concreteness we shall assume $(3+1)$ dimensions, even
though the generalization to other dimensions is straightforward.
Let $z=z_w(t)$ be the trajectory of the 
moving boundary, assumed timelike but otherwise arbitrary. 
In the semi-infinite region $z>z_w(t),$ with
$t,$ $x,$ $y$ arbitrary, the wave equation 
\ba
A_{\mu ,\nu \nu }=0
\ea
describes the forward time evolution.
There remains, however, some residual gauge freedom, which may be 
completely characterized by the scalar solutions to the wave equation
\ba
\Lambda _{,\nu \nu }=0.
\ea
It follows that the gauge transformation 
\ba
A_\mu \to A_\mu ^\prime =A_\mu -\Lambda _{,\mu }
\ea
preserves the Lorentz gauge condition
\ba
A_{\mu ,\mu }=0
\label{lgc-a}
\ea
and describes the same classical field configuration. 
The residual gauge freedom is quite limited. 
Once initial boundary data have been specified (i.e., on
a past Cauchy surface),
no ambiguity remains as to the evolution of the vector potential 
forward in time. Once $A_\mu $ and its normal
derivative have been specified on a Cauchy surface, the future
propagation is completely unambiguous, because the initial data
for $\Lambda $ may be propagated forward in time from the same
Cauchy surface. For a spatially semi-infinite region, however, pure gauge
modes may propagate in from the timelike boundary, as indicated
in Fig.~\ref{Fig:ads}. In order to fix the time evolution completely, some sort
of gauge condition on this timelike boundary is required to specify
what longitudinal modes, if any, propagate in and how longitudinal
modes are reflected off the boundary. 

We first consider a boundary moving at a constant velocity, which may 
be taken to be situated at $z=0.$ ${\bf E}$ and ${\bf B}$ vanish inside
the conducting wall. (We assume that the conductor has no 
magnetic flux frozen in.) 
It follows that ${\bf E}_\parallel $ and ${\bf B}_\perp $ vanish at
the boundary, 
although ${\bf E}_\perp $ and ${\bf B}_\parallel $ need not vanish
there, because 
these may be generated by surface charge and current density, respectively,
on the boundary.  
We fix the gauge inside the conductor by setting 
$\underline{\bf A}=(A_0,{\bf A})=0$ there. 
It follows that ${\bf A}_\parallel $ and $A_0$---or, in 
other words, all the components parallel to the 
boundary---must vanish. A jump in $A_\perp ,$ however, is not excluded.
Such a jump can be generated from a gauge transformation of the 
${\bf E}={\bf B}=0$ configuration, for example one where $\Lambda $
vanishes inside the conductor and rises linearly with distance from
it on the exterior.  

In the bulk $z>0,$ for fixed wavenumber 
$\underline{\bf k}=({\bf k}_\parallel ,k_\perp ),$ 
there are three modes consistent with the Lorentz gauge condition in 
eqn.~(\ref{lgc-a}):
two physical modes
\ba
\hat {\bf e}^\mu_{(1)}(\hat {\bf k})~
 \exp [+i\underline{\bf k}\cdot \underline{\bf x}],\quad
\hat {\bf e}^\mu_{(2)}(\hat {\bf k})~
 \exp [+i\underline{\bf k}\cdot \underline{\bf x}]
\ea
and a longitudinal mode 
\ba
\hat {\bf e}^\mu_{(L)}(\hat {\bf k})~
 \exp [+i\underline{\bf k}\cdot \underline{\bf x}] ,
\ea
where 
\ba 
\hat {\bf e}^\mu_{(1)}(\hat {\bf k})= {
\hat {\bf e}^\mu _{x}-\hat {\bf k}^\mu 
 (\hat {\bf k}\cdot \hat {\bf e}_{x})\over 
 \sqrt{1-(\hat {\bf k}\cdot \hat {\bf e}_{x})^2} 
},\quad
\hat {\bf e}^\mu_{(2)}(\hat {\bf k})= {
\hat {\bf e}^\mu _{y}-\hat {\bf k}^\mu 
 (\hat {\bf k}\cdot \hat {\bf e}_{y})\over  
 \sqrt{1-(\hat {\bf k}\cdot \hat {\bf e}_{y})^2}
},\quad 
\hat {\bf e}^\mu_{(L)}(\hat {\bf k})
={1\over \sqrt{2}}\Bigl( \hat {\bf e}^\mu _{t}+\hat {\bf k}^\mu \Bigr)
\ea
and 
$\hat {\bf k}=\underline{\bf k}/\vert \underline{\bf k}\vert ,$ 
$\hat {\bf e}^\mu _{t}=(1,0,0,0).$
A fourth polarization 
\ba
\hat {\bf e}_{(F)}=
{1\over \sqrt{2}}
(\hat {\bf e}_{t}-\hat {\bf k}),
\ea
which we shall call the forbidden polarization, is excluded by
the Lorentz gauge condition. 

We have already specified boundary conditions for the components of
the 4-vector 
potential parallel to the conducting boundary, in the present special case
$A_t,$ $A_x,$ and $A_y.$ These components satisfy Dirichlet boundary
conditions. 
It remains to determine the boundary condition for the normal component 
$A_\perp,$ $A_z$ for the special case here. It is apparent
that for incoming longitudinal 
modes to scatter off the conductor into outgoing longitudinal modes,
without creating  
any of the forbidden fourth polarization, a Neumann boundary condition
is required,  
in other words, $A_{z,z}=0.$ More generally, 
\ba
\hat {\bf t}\cdot \underline{\bf A}=0,\quad 
\nabla _{\hat {\bf n}}(\hat {\bf n}\cdot \underline{\bf A})=0,
\ea
where $\hat {\bf t}$ is an arbitrary vector tangent to the conducting
surface and $\hat {\bf n}$ is the normal vector. 

This nice separation into physical and pure gauge (longitudinal) modes 
is not preserved under Lorentz boosts. While a pure gauge mode always
transforms into another pure gauge mode, a physical mode transforms into
a physical mode with some, in general non-vanishing, admixture of the 
longitudinal (pure gauge) mode. It is not possible for observers moving
relative to each other to agree on a common notion for the absence of
a longitudinal component.

For the case of a stationary conducting boundary, it is possible to
remove this ambiguity by further fixing the gauge to Coloumb gauge,
where $A_0=0,$ which is equivalent to postulating the absence of
longitudinal photons for the preferred stationary observers. However,
when the velocity of the boundary is not constant, physical photons
reflected from the boundary will, upon being diffracted back, acquire
a longitudinal component from the point of view of an observer at rest
with respect to the boundary at a later instant. Consequently, it is
necessary to admit longitudinal modes when considering curved or
accelerating boundaries.  

\subsection{Reflection off a planar perfect gravitational mirror}

We now consider linearized gravity for a metric perturbation 
$h_{\mu \nu}$ of Minkoswki space having the background metric
$\eta_{\mu \nu}.$
Under a gauge transformation generated by the displacement field
$\xi ^\mu ({\bf \underline{x}}),$ the metric perturbation $h_{\mu \nu },$
defined so that $g_{\mu \nu }=\eta _{\mu \nu }+h_{\mu \nu },$ transforms as 
\ba 
h_{\mu \nu }\to h_{\mu \nu }^\prime =h_{\mu \nu }-\xi _{\mu ,\nu }
-\xi _{\nu ,\mu }.
\ea
It is convenient to define the trace-reversed metric perturbation
\ba
\bar h_{\mu \nu }=h_{\mu \nu }
-{1\over 2}\eta _{\mu \nu }\eta ^{\rho \sigma}h_{\rho \sigma}.
\ea
[In our tensor calculus, valid only to first order, indices are
raised and lowered with the unperturbed metric
$\eta _{\mu \nu }.$]
We impose the Lorentz gauge condition
\ba
\bar h_{\mu \nu ,\nu }=0.
\label{lgc}
\ea
It follows that a displacement field satisfying 
\ba
\Box ~\xi _\mu =0
\ea 
preserves the gauge condition in eqn.~(\ref{lgc}), and that the equation
of motion for the metric perturbation becomes
\ba
\Box ~\bar h_{\mu \nu }=-(16\pi G)~T_{\mu \nu },
\ea
which for a vacuum stress-energy tensor becomes
\ba
\Box ~\bar h_{\mu \nu }=0.
\ea

If we consider plane waves with the spatial-temporal dependence
\ba
\exp [i{\bf \underline{k}}\cdot {\bf \underline{x}}] = 
\exp [i{\bf {k}}\cdot {\bf {x}}-i\omega t],
\ea
we find that
there are a total of ten possible polarizations for $\bar h_{\mu\nu}$: 
two physical polarizations, four longitudinal (pure gauge) polarizations, 
and four forbidden polarizations, excluded by the Lorentz gauge
condition. Explicitly, these are as follows. The two physical
polarizations are
\ba
{\bf \hat{\hat{e}}}_{(+)} &=& {1\over \sqrt{2}}\Bigl(
 {\bf \hat e}_{(1)}\otimes{\bf \hat e}_{(1)}
-{\bf \hat e}_{(2)}\otimes{\bf \hat e}_{(2)} 
\Bigr) ,\cr
{\bf \hat{\hat{e}}}_{(X)}&=& {1\over \sqrt{2}}\Bigl(
 {\bf \hat e}_{(1)}\otimes{\bf \hat e}_{(2)}
+{\bf \hat e}_{(2)}\otimes{\bf \hat e}_{(1)}
\Bigr) .
\ea
The four longitudinal (pure gauge) polarizations are
\ba
{\bf \hat{\hat{e}}}_{(gF)}&=& {1\over \sqrt{2}}\Bigl(
 {\bf \hat e}_{(1)}\otimes{\bf \hat e}_{(1)}
+{\bf \hat e}_{(2)}\otimes{\bf \hat e}_{(2)}
\Bigr) ,\cr
{\bf \hat{\hat{e}}}_{(gL)}&=& {\bf \hat e}_{(L)}
 \otimes{\bf \hat e}_{(L)},\cr
{\bf \hat{\hat{e}}}_{(g1)}&=& {1\over \sqrt{2}}\Bigl(
 {\bf \hat e}_{(1)}\otimes{\bf \hat e}_{(L)}
+{\bf \hat e}_{(L)}\otimes{\bf \hat e}_{(1)} 
\Bigr),\cr  
{\bf \hat{\hat{e}}}_{(g2)}&=& {1\over \sqrt{2}}\Bigl(
 {\bf \hat e}_{(2)}\otimes{\bf \hat e}_{(L)}
+{\bf \hat e}_{(L)}\otimes{\bf \hat e}_{(2)}
\Bigr),
\ea
and the four forbidden polarizations are
\ba
{\bf \hat{\hat{e}}}_{(fF)}&=& {\bf \hat e}_{(F)}
 \otimes{\bf \hat e}_{(F)},\cr 
{\bf \hat{\hat{e}}}_{(fL)}&=& {\bf \underline{\underline{\bf g}}},\cr
{\bf \hat{\hat{e}}}_{(f1)}&=& {1\over \sqrt{2}}\Bigl(
 {\bf \hat e}_{(1)}\otimes{\bf \hat e}_{(F)}
+{\bf \hat e}_{(F)}\otimes{\bf \hat e}_{(1)}
\Bigr) ,\cr
{\bf \hat{\hat{e}}}_{(f2)}&=& {1\over \sqrt{2}}\Bigl(
 {\bf \hat e}_{(2)}\otimes{\bf \hat e}_{(F)}
+{\bf \hat e}_{(F)}\otimes{\bf \hat e}_{(2)}
\Bigr) .
\ea
Here the double hats denote unit vectors for a second-rank tensor.

As a boundary condition, let us first consider an ideal gravitational mirror
in the form of a brane with vanishing stress-energy density localized on
the brane and a $Z_2$ orbifold symmetry across the brane. We take the geometry
of the unperturbed brane to be flat. This idealized invisible brane is
the short distance limit of any sort of actual brane, because for any actual
nonsingular brane, there is a finite length scale characterizing its vacuum
curvature and that inducing by the matter on it. 

In the usual Cartesian coordinates, we obtain the following boundary
conditions for a  
normally incident gravitational wave. Let ${\underline{\bf t}}_1$ and 
${\underline{\bf t}}_2$ be arbitrary vectors tangent to the brane. One has
the boundary condition
\ba 
{\partial \over \partial n}
\biggl(
{\underline{\bf t}}_1 \cdot 
\underline {\underline {\bf h}}
\cdot {\underline{\bf t}}_2 \biggr) =0 ,
\ea
as derived from the Israel matching conditions,
where $\partial /\partial n$ denotes the inward normal derivative. We
deduce the boundary conditions for the remaining components by
considering the scattering of incoming longitudinal (pure gauge)
graviton modes. It is necessary that these scatter 
exclusively into outgoing longitudinal modes. It is evident that the
boundary conditions 
\ba 
{\underline{\bf n}} \cdot 
\underline {\underline {\bf h}}
\cdot {\underline{\bf t}}_1 &=&0,\cr
{\partial \over \partial n}
\biggl( 
{\underline{\bf n}}\cdot 
\underline {\underline {\bf h}}
\cdot {\underline{\bf n}}
\biggr) &=&0
\ea 
for the other components satisfy this requirement, for normally as well as 
for obliquely incident waves. 

\section{Lorentz Gauge in a Curved Background Spacetime}
\label{Lorentz Gauge in a Curved Background Spacetime}

In a curved spacetime, under the gauge transformation generated by an
infinitesimal (linearized) 
displacement field $\xi^{a}$, the linearized 
metric perturbation transforms according to
\ba
h_{ab} \to  h^{\prime}_{ab}=h_{ab}-{\cal L}_{\xi}g^{(0)}_{ab}
=h_{ab}-\xi_{a;b}-\xi_{b;a}.
\ea
The trace modified metric perturbation consequently
\ba
{\bar h}_{ab}
=h_{ab}-{1\over 2}g^{(0)}_{ab} \Bigl( g^{(0)~cd}~h_{cd} \Bigr)
\label{eqn:hbar}
\ea
transforms as
\ba
{\bar h}_{ab} \to  {\bar h}^{\prime}_{ab}
={\bar h}_{ab}-\xi_{a;b}-\xi_{b;a}
+g^{(0)}_{ab}{\xi^{c}}_{;c}.
\ea
For gauge transformations that preserve the Lorentz 
gauge condition eqn.~(\ref{lgc-z}), $\delta \bar h_{ab;b}=0,$
which implies that $\xi^{a}$ satisfies
\ba
\Box \xi_{a} +{\xi^{b}}_{;ab} 
-{\xi^{b}}_{;ba} =
\Box \xi_{a} +R _{ab}^{(0)}~\xi^{b}=0.
\label{eqn:gauge}
\ea
where
$\Box =g^{ab}~\nabla_{a}\nabla_{b}.$
The evolution equation for $\bar h_{ab}$ 
is [see Appendix \ref{appendix:evo:eqn:h} for a derivation]
\ba
\Box \bar h_{ab}
-\left[
g^{(0)}_{bc} \bar h_{da}
+g^{(0)}_{ac} \bar h_{db}
+g^{(0)}_{ab} \bar h_{cd}
-g^{(0)}_{cd} \bar h_{ab}
\right] R^{(0)}_{cd}
+2R^{(0)}_{acbd}~\bar h_{cd}
=-(16\pi G)~T^{(1)}_{ab}.
\label{eqn:h:wave:source}
\ea
When the background is
maximally symmetric, a number of simplifications result.
For a (d+1)-dimensional maximally symmetric background spacetime
\ba
R_{abcd}^{(0)} ={R^{(0)}\over {d(d+1)}}\left( g_{ac}^{(0)}~g_{bd}^{(0)} 
-g_{ad}^{(0)}~g_{bc}^{(0)} \right),
\quad
R_{ab}^{(0)} ={R^{(0)}\over (d+1)} g_{ab}^{(0)},
\ea
so that eqn.~(\ref{eqn:gauge}) becomes
\ba
\left(\Box +{R^{(0)}\over {d+1}}\right)\xi^{a} =0,
\label{eqn:gauge:wave:MaxSymm}
\ea
and eqn.~(\ref{eqn:h:wave:source}) becomes
\ba
\Box \bar h_{ab} 
-{R^{(0)}\over {d+1}} \left[
-(d-1)\bar h_{ab} +g^{(0)}_{ab}\bar h_{cc} \right]
=-(16\pi G)~T^{(1)}_{ab}.
\label{eqn:h:wave:MaxSymm}
\ea
For $AdS^5$ presented using the Randall-Sundrum line element
\ba
ds^2 =\frac{\ell^2}{z^2}\left[ dz^2 -dt^2 +d{\bf x}^2\right] ,
\ea
we find that 
\ba
R_{abcd}=-\frac{1}{\ell^2}\left( g_{ac}~g_{bd} 
-g_{ad}~g_{bc}\right), \quad
R_{ab}=-\frac{4}{\ell^2}g_{ab},\quad
R=-\frac{4\cdot 5 }{\ell^2},
\ea
and $T_{ab}=-\Lambda g_{ab}$ where
$\Lambda =-3/(4G\ell^2 ),$
which when substituted in eqn.~(\ref{eqn:h:wave:MaxSymm}) yields
\ba
\Box \bar h_{ab}=0.
\label{eqn:h:wave:AdS5}
\ea

In (4+1) dimensions the graviton field $h_{a b}$ has
$(5\cdot 6)/2=15$ independent components or polarizations. 
Under the Lorentz gauge condition 
five of these are rendered forbidden polarizations. The residual
gauge freedom may be characterized completely by the solutions to
the homogeneous 
vector wave equation eqn.~(\ref{eqn:gauge}). Here
$\xi ^a ({\bf \underline{x}})$ is simply an
infinitesimal displacement field generating a coordinate 
reparameterization. This residual gauge freedom corresponds to 
five longitudinal (pure gauge) graviton polarizations, leaving
a total of five ``physical" polarizations. Again, observers moving
at a constant velocity with respect to each other cannot
agree on a common notation of a purely physical mode without any
admixture of longitudinal polarizations. 

\def\x{{\bf \underline{x}}}

Before proceeding to a further analysis of the residual gauge
freedom in Lorentz gauge, let us first, for comparison, review the counting of
allowed polarization states and residual gauge freedom in the 
Randall-Sundrum gauge (see Ref.~\cite{rs}, \cite{gt1}), where 
$h_{55}=h_{5\mu }={h^\mu }_\mu =h_{\mu \nu ,\nu }=0.$ There are
five remaining polarizations, all
of which are physical. Fixing the gauge completely in this way
in general displaces the brane from its non-perturbed position
at $z=1.$ Consequently, in order to account for all the physical
degrees of freedom, it is also necessary to include an additional 
scalar field, which we shall denote 
$\xi _{brane \perp }({\bf x }),$ localized on the brane, indicating a normal
displacement of the brane with respect to the bulk. [In our notation,
the vector ${\bf x }$ denotes a position on the brane whereas $\x $
denotes a position in the bulk.] 
In Lorentz gauge this last degree of freedom is absent because
the corresponding physical degree of freedom is instead described
by one of the five longitudinal graviton polarizations. If 
we dispense with the normal displacement of the brane with respect
to the bulk geometry, then the brane is fixed to the surface
$z=z_{(b)}(t),$ with $t,$ $x_1,$ $x_2$ and $x_3$ arbitrary, no
matter what the perturbation is. 

Consider now a gauge transformation
generated by the displacement field $\xi ^a (\x ).$ For this
to be pure gauge, we must displace the brane about its unperturbed
position $z=z_{(b)}(t)$ according to the normal displacement field
$\xi _{brane \perp }({\bf x })=n _a \xi ^a (\x ).$ However, if instead
we fail to compensate for the change in position of the brane
by setting $\xi _{brane \perp }({\bf x })$ equal to zero, what would
be a pure gauge transformation corresponds to a displacement of
the brane trajectory relative to the bulk geometry. The remaining
four longitudinal polarizations allowed in Lorentz gauge
correspond to reparameterizations of the brane. 

The Lorentz gauge condition in the bulk does not in any way
fix the gauge on the brane. A gauge transformation corresponding
to a reparameterization of the brane is effected by a flux of
incoming longitudinal gravitons emanating from the Cauchy
surface and reflecting off the brane. More specifically,
let the field $\xi _\parallel (\x )$ indicate an infinitesimal
reparameterization of the brane. The longitudinal gravitons
implementing the desired gauge transformation are generated
by choosing  $\xi (\underbar \x )$ on the initial Cauchy surface
such that 
$\xi _\parallel (\x )$ on the boundary is reproduced
and 
$\xi _\perp (\x )$ vanishes there.
Because we do not include an additional degree of freedom
corresponding to normal displacements of the brane, from
the view of an observer on the brane, the longitudinal
modes in the bulk giving vanishing 
$\xi _\parallel (\x )$ but non-vanishing $\xi _\perp (\x ) $
on the brane are not ``pure gauge''. These longitudinal
modes take the place of the field localized to the brane
$\xi _{brane \perp }(\x )$ present in the Randall-Sundrum gauge.

\section{Compatibility of Sources on the 
Boundary with the Gauge Condition}
\label{Compatibility of Sources on the 
Boundary with the Gauge Condition}

In this section we investigate the compatibility of boundary 
conditions on the timelike boundary (i.e., the brane) with
the Lorentz gauge condition. First, as a sort of warm-up
exercise, we consider a flat bulk with a planar boundary,
which acts as a perfect gravitational mirror. Then we consider
the modifications necessary to take into account extrinsic 
curvature (i.e., nonvanishing stress-energy on the brane at zeroth
order) and the spacetime curvature of the bulk at zeroth
order for perfect AdS.

\subsection{Special case---the planar perfect gravitational mirror 
embedded in a Minkowski bulk revisited}

We first consider the following boundary conditions for a 
linearized source on the gravitational mirror boundary
(here, in this special case,
the bulk is Minkowski space and so is the unperturbed
brane)
\ba
\bar h_{AB,N}&=&-\kappa^2_{(5)}{\cal T}_{AB},\label{aaa}\\
\bar h_{AN}&=&0,\label{aab}\\
\bar h_{NN,N}&=&0.\label{aac}
\ea
Here the indices $A,B,\ldots $ indicate directions tangential to the brane,
$N$ the normal direction, and $a,b,\ldots $ all five directions.
${\cal T}_{AB}$ denotes here the linearized stress-energy perturbation
on the brane.
Eqn.~(\ref{aaa}) is the Israel matching condition and eqns.~(\ref{aab})
and (\ref{aac}) are supplementary gauge conditions on the boundary.
We show that the Lorentz gauge condition is 
subsequently satisfied if and only if it is
satisfied on the initial Cauchy surface and ${\cal T}_{AB,B}=0$
as well. 

To this end, we consider the vector field 
\ba
V_a=\bar h_{ab,b},
\label{v_field}
\ea
assumed to vanish and have vanishing normal derivative
on the initial Cauchy surface.
If each component of this 4-vector field can be shown to satisfy either
Dirichlet or Neumann homogeneous boundary conditions on the timelike
boundary (i.e., the brane), it follows that the Lorentz gauge condition
holds everywhere. Physically, this means that wave packets obeying the 
gauge condition reflect into wave packets obeying the gauge condition
and that any inhomogeneous source on the timelike boundary does not
emit any waves violating the Lorentz gauge condition.

On the boundary, the relation 
\ba
V_N=\bar h_{NN,N}+\bar h_{AN,A}=0
\ea
holds
trivially because of the supplementary gauge 
conditions (\ref{aab}) and (\ref{aac}). 
Similarly, the normal derivative of the tangential components is
given by 
\ba
V_{A,N}
=\bar h_{AN,NN}+\bar h_{AB,BN}
=-\bar h_{AN,BB}-\kappa^2_{(5)}{\cal T}_{AB,B}
=-\kappa^2_{(5)}{\cal T}_{AB,B}.
\ea
Consequently, no Lorentz gauge condition violating waves emanate from
the boundary if and only if 
\ba
{\cal T}_{AB,B}=0,
\label{em-cons}
\ea
in other words, if ${\cal T}_{AB}$ is a conserved tensor field on the brane.

\subsection{The general case---bulks and branes curved at zeroth order}

In the previous section we were able to show in a few lines, relying
on the wave equation in the bulk and stress-energy conservation on
the brane, that (\ref{aab}) and (\ref{aac}) are the correct auxilliary
boundary conditions in Lorentz gauge if one assumes that both the 
brane and the bulk are flat at zeroth order. 
In this section we essentially repeat the derivation of the previous section
for the case where the boundary and 
the bulk geometry are not flat, including modifications necessary
to account for the non-vanishing curvature. When the bulk is
curved, the condition $\bar h_{AN}=0$ no longer necessarily
implies that $\bar h_{AN;A}=0$ on the boundary. Consequently, it is
necessary to modify  
eqn.~(\ref{aac}) to contain non-derivative contributions of the metric
perturbation. 

To treat the curved case, we first fix a point 
${\bf x}$ on the boundary $\Sigma .$
Then we coordinatize a neighborhood of ${\bf x}$ on 
$\Sigma $ using the coordinates $x_A$
chosen to be Fermi normal coordinates---that is, 
using exponential map from the tangent space of 
${\bf x}$ on $\Sigma $ to $\Sigma .$ By choosing
Fermi normal coordinates we set all the Christoffel
symbols of the metric connection on $\Sigma $ to zero at ${\bf x},$
although the first and higher partial derivatives of the Christoffel symbols
do not in general vanish there. These coordinates may be continued
off the boundary by means of the Gaussian normal prescription---that is, 
the coordinate $n$ in the $N$ direction vanishes on 
$\Sigma $ and the coordinates on $\Sigma $ are continued along
initially normal geodesics, $n$ indicating 
physical distance along these geodesics from $\Sigma .$ In this way, 
at ${\bf x}$ we obtain the following connection coefficients
for the unit vectors $\hat e_A=\partial /\partial x_A$ and
$\hat e_N=\partial /\partial n:$ 
\ba 
\nabla _N\hat e_N&=&0,\cr
\nabla _A\hat e_N&=&+K_{AB}~\hat e_B,\cr
\nabla _N\hat e_A&=&+K_{AB}~\hat e_B,\cr
\nabla _A\hat e_B&=&-K_{AB}~\hat e_N,
\ea
where the tensor $K_{AB}$ is the extrinsic curvature 
and $\nabla $ denotes the bulk covariant derivative induced by the bulk
metric. These equations may be explained as follows. The
first equation simply defines the continuation of $N$
off the brane. The second equation consists merely a restatement
of 
the definition of the extrinsic
curvature. The third equation follows from the vanishing of the Lie
derivative of $N$ along the brane (i.e., $[A,N]=0).$ 
Finally, the fourth equation
follows because $N$ is orthogonal to $\Sigma.$
Away from ${\bf x},$ the last equation would be modified to
\ba
\nabla _A\hat e_B&=&\tilde \Gamma ^C_{AB}~\hat e_C
-K_{AB}~\hat e_N,
\ea
where $\tilde \Gamma ^C_{AB}$ denotes the connection coefficients
of the covariant derivative on
$\Sigma $ with respect to the metric induced on $\Sigma $ by the bulk metric.
For future reference, we
also give the following second covariant derivatives along the brane
\ba
\nabla _D\nabla _C~\hat e_N
&=&
+K_{CA;D}~\hat e_A -K_{CA}~K_{DA}~\hat e_N,\cr
\nabla _D\nabla _C~\hat e_B
&=&
\left( 
\tilde \Gamma ^{A}_{CB;D}
-K_{CB}~K_{DA}\right)\hat e_A
-K_{CB;D}~\hat e_N.
\ea

\subsubsection{The electromagnetic case}

For the electromagnetic case we calculate on the boundary at ${\bf x}$
\ba
A_{a;a}&=&A_{A;A}+A_{N;N}\cr
&=&A_{N;N}
+\nabla _A\left[ A_B^{(c)}~\hat e_B
+A_N^{(c)}~\hat e_N\right] \cdot \hat e_A\cr
&=&A_{N;N} +{\partial A_B^{(c)}\over \partial x_B}+K_{AA} A_N^{(c)}.
\label{fff}
\ea
Here the superscript $(c)$ denotes that the component in question is to be 
regarded as a scalar quantity (i.e., differentiated using
ordinary partial rather than covariant differentiation). From our
boundary condition 
$A_B=0$ (i.e., $\underline{t}\cdot 
\underline{A}=0$), it follows that the second term vanishes.
However, to ensure the reflections and sources from $\Sigma $ do not
violate the Lorentz gauge condition, it is necessary to modify the 
supplementary boundary condition to
\ba 
A_{N;N} +K_{AA}~A_N=0,
\ea
so that eqn.~(\ref{fff}) is set to zero.

\subsubsection{Linearized gravitational perturbations}

For linearized gravity on a curved boundary
only the following boundary condition stays unaltered
\ba
\bar h_{AN}=0.
\label{eqn:hAN}
\ea
This is because adding any derivative term would spoil the short 
distance limit. Consequently, we retain the boundary condition
in eqn.~(\ref{eqn:hAN}) and next find the modifications to eqn.~(\ref{aac})
required for the curved case.
We find that 
\ba
V_{N}
&=&\bar h_{NN;N} +\bar h_{NA;A} \cr
&=&\bar h_{NN;N} \cr
&&+\nabla_{A}\Bigl[
 \bar h^{(c)}_{NN}~(\hat e_{N}\otimes\hat e_{N})
+\bar h^{(c)}_{ND}~(\hat e_{N}\otimes\hat e_{D}) \cr
&& \quad+\bar h^{(c)}_{CN}~(\hat e_{C}\otimes\hat e_{N})
+\bar h^{(c)}_{CD}~(\hat e_{C}\otimes\hat e_{D})
\Bigr] \cdot (\hat e_{N}\otimes\hat e_{A}) \cr
&=&\bar h_{NN;N} +\bar h_{NN}~K_{AA} -\bar h_{BA}~K_{AB}.
\ea
The above
implies that the boundary condition valid for the case with a flat
brane and a flat bulk, $\bar h_{NN,N}=0$ [eqn.~(\ref{aac})], must be
modified to
\ba
\bar h_{NN;N} +K_{AA}~\bar h_{NN} -K_{AB}~\bar h_{AB} =0
\label{eee}
\ea
for the general case. At this point that the boundary condition
eqn.~(\ref{eqn:hAN}) remains unaltered is just an inspired guess,
vindicated in the calculations that follow.

We now compute the normal derivative of the tangential component of the
gauge condition,
\ba
V_{A;N} &=&\bar h_{AN;NN} +\bar h_{AB;BN}\cr
        &=&-\bar h_{AN;BB} +\bar h_{AB;NB}\cr
        &&+(\bar h_{AN;NN}+\bar h_{AN;BB})\cr
        &&+(\bar h_{AB;BN}-\bar h_{AB;NB}).
\ea
The second line of the second expression can be simplified by using the
wave equation in eqn.~(\ref{eqn:h:wave:source}) and the Gauss-Codacci
relations in eqn.~(\ref{eqn:Gauss:Codacci})
\ba
\bar h_{AN;NN}+\bar h_{AN;BB}
=\Box ~\bar h_{AN} 
=-2R_{NDAC}~\bar h_{CD} 
=-2\left( K_{DA;C}^{(0)} -K_{DC;A}^{(0)}\right) \bar h_{CD},
\ea
and the third line simplifies using 
\ba
\bar h_{AB;BN}-\bar h_{AB;NB} 
=R_{BCNB}~\bar h_{AC} +R_{ACNB}~\bar h_{CB}
=\left( K_{BA;C} -K_{BC;A}\right)\bar h_{CB},
\ea
so that
\ba
V_{A;N} 
&=&-\bar h_{AN;BB} +\bar h_{AB;NB}
-\left( K_{BA;C}^{(0)} -K_{BC;A}^{(0)}\right) \bar h_{CB}.
\ea
We now evaluate $\bar h_{AN;BB}$.
From the definition of the Laplacian operator
\ba
\Box 
&=&\left(N^{\mu}N^{\nu} +\sum_{A}A^{\mu}A^{\nu} \right)
 \nabla_{\mu}\nabla_{\nu} \cr
&=&
\left(N^{\mu}\nabla_{\mu} \right) 
 \left(N^{\nu}\nabla_{\nu} \right)
-\left(N^{\mu} \nabla_{\mu}N^{\nu} \right)\nabla_{\nu} 
+
\sum_{A} \left[
\left(A^{\mu}\nabla_{\mu}\right) 
 \left(A^{\nu}\nabla_{\nu} \right)
-\left(A^{\mu} \nabla_{\mu}A^{\nu} \right)\nabla_{\nu} \right] \cr
&=&
\nabla_{N}\nabla_{N}
+\sum_{A} \left[\nabla_{A}\nabla_{A} +K_{AA}^{(0)}\nabla_{N}\right],
\ea
it follows that
\ba
\bar h_{AN;BB}=
\left[\nabla_{B}\nabla_{B} +K_{BB}^{(0)}\nabla_{N}\right]\bar h_{AN}.
\ea
To simplify $\bar h_{AN;BB}$ exploiting the fact that $\bar h_{AN}=0$
on the brane, we compute
\ba
\nabla _D\nabla _C \bar h_{AN}
&=&
\nabla _D\nabla _C ~
\Bigl[
 \bar h^{(c)}_{NN}~(\hat e_{N}\otimes \hat e_{N})
+\bar h^{(c)}_{NE}~(\hat e_{N}\otimes \hat e_{E}) \cr 
&&\quad +\bar h^{(c)}_{EN}~(\hat e_{E}\otimes\hat e_{N})
+\bar h^{(c)}_{EF}~(\hat e_{E}\otimes \hat e_{F})
\Bigr] \cdot (\hat e_{A}\otimes \hat e_{N})\cr
&=& \nabla _D\bar h^{(c)}_{NN}~(\nabla _C\hat e_{N})\cdot \hat e_{A}
+\nabla _C\bar h^{(c)}_{NN}~(\nabla _D\hat e_{N})\cdot \hat e_{A}
+\bar h^{(c)}_{NN}~(\nabla _D\nabla _C\hat e_{N})\cdot \hat e_{A} \cr
&&+\bar h^{(c)}_{AF}~(\nabla _D\nabla _C\hat e_{F})\cdot \hat e_{N}
+\nabla _D\bar h^{(c)}_{AF}~(\nabla _C\hat e_{F})\cdot \hat e_{N}
+\nabla _C\bar h^{(c)}_{AF}~(\nabla _D\hat e_{F})\cdot \hat e_{N} \cr
&=&
K^{(0)}_{AC}~\bar h_{NN;D}
+K^{(0)}_{AD}~\bar h_{NN;C}
+K^{(0)}_{AC;D}~\bar h_{NN}\cr
&&
-K^{(0)}_{FC;D}~\bar h_{AF}
-K^{(0)}_{CF}~\bar h_{AF;D}
-K^{(0)}_{DF}~\bar h_{AF;C}.
\ea
Contracting $C$ and $D,$ we obtain
\ba
\nabla _B\nabla _B \bar h_{AN}&=&
2 K^{(0)}_{AB}~\bar h_{NN;B}
+K^{(0)}_{AB;B}~\bar h_{NN}
-K^{(0)}_{CB;B}~\bar h_{AC}
-2K^{(0)}_{BC}~\bar h_{AC;B}.
\ea
Consequently,
\ba
V_{A;N}
&=&
\bar h_{AB;NB} -\bar h_{AN;BB} 
-\left( K_{BA;C}^{(0)} -K_{BC;A}^{(0)}\right) \bar h_{CB}\cr
&=& 
\bar h_{AB;NB}
-2K^{(0)}_{AB}~\bar h_{NN;B} 
+2K^{(0)}_{BC}~\bar h_{AB;C} 
-K^{(0)}~\bar h_{AN;N} \cr
&&
-K^{(0)}_{AB;B}~\bar h_{NN}
+K^{(0)}_{BC;C}~\bar h_{AB} 
-\left( K_{BA;C}^{(0)} -K_{BC;A}^{(0)}\right) \bar h_{CB}.
\label{eqn:va;n}
\ea
It remains to be shown using the Israel matching condition and 
stress-energy conservation on the brane that the above quantity
vanishes, so that the corresponding boundary condition on the 
brane is homogeneous.

\section{Stress-energy Conservation on the Brane at First Order}
\label{Stress-energy Conservation on the Brane at First Order}

\begin{figure}[t]
\setlength{\unitlength}{1cm}
\begin{center}
\begin{minipage}[t]{7.cm}
\begin{picture}(7.,7.)
\centerline 
{\hbox{\psfig{file=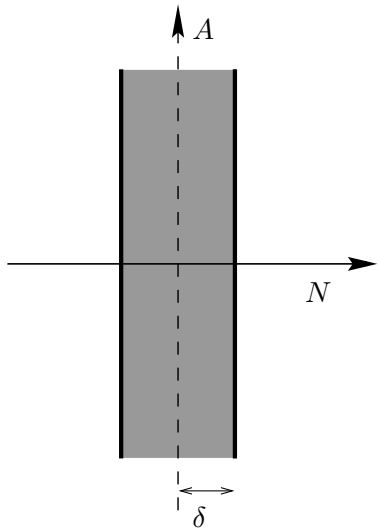,height=7.cm,width=5.cm}}}
\put(-3.5,0){$\delta $}
\put(-2.,3){$N $}
\put(-3.5,6.5){$A$}
\end{picture}\par
\end{minipage}
\end{center}
\caption{\small \baselineskip=4pt { 
{\bf Schematic diagram of the $Z_2$ symmetric distribution of
brane stress-energy.} 
A singular distribution of stress-energy on
the brane is most easily regarded as the limit of a 
non-singular distribution of stress-energy of support
confined to within a shell of thickness $2\delta $ 
in the limit $\delta \to 0+.$ We
derive the boundary conditions applicable for an infinitely thin
singular brane by taking the  $\delta \to 0+$ limit of a  nonsingular
thick brane of thickness $\delta .$ The (dashed) line into the
middle represents the $Z_2$ reflection orbifold symmetry.  The
projected stress-energy ${\cal T}_{AB}$ (always tangent to the brane) is
distributed over the indicated shaded region of thickness $2\delta .$
}}
\label{Fig:Z2}
\end{figure}

In this section we consider how to express 
stress-energy conservation for a singular
distribution of co-dimension one of stress-energy localized
on the boundary brane about which a $Z_2$ symmetry has been 
imposed. The simplest way to proceed, which does not require 
any additional principles for how to treat the singular
distribution of boundary matter, is to consider in the 
limit $\delta \to 0\,+$ a symmetric
distribution of non-singular $Z_2$ symmetric matter of a finite
small thickness $2\delta .$ The
bulk spacetime is reflected about the hypersurface at the 
center of this boundary stress-energy by means of the 
$Z_2$ symmetry, as indicated in Fig.~\ref{Fig:Z2}. 
The four-dimensional projected stress-energy tensor          
on the brane ${\cal T}_{AB}$ is obtained by integrating over the coordinate
normal to the brane, denoted by $N,$ whose units
correspond to physical distance. One has
\ba
{\cal T}_{AB}({\bf x})=\int _{-\delta }^{+\delta }dN~
T_{AB}({\bf x},N),
\ea
where $T_{AB}$ is the five-dimensional stress-energy tensor.
For finite $\delta $ this projection procedure is not entirely
satisfactory because of ambiguities in parallel transport. 
However as $\delta \to 0\,+$ these difficulties                   
disappear.

Using ordinary derivatives, we may write
\ba
\int _{-\delta }^{+\delta }dN~
\left(
T_{AB,B} + T_{AN,N}
\right)
&=&{\cal T}_{AB,B}+T_{AN}(N=+\delta )-T_{AN}(N=-\delta )\cr 
&=&{\cal T}_{AB,B}+2T_{AN}(N=+\delta ).
\ea
Here we may choose, for the purpose of this calculation, Gaussian
normal coordinates, so that $N=0$ corresponds to the hypersurface
$\Sigma $ of $Z_2$ symmetry, and $N$ is the signed normal
physical distance from $\Sigma.$ In the neighborhood of a
given point ${\bf x}\in \Sigma ,$ we may choose the transverse
coordinates, labelled $A,B,\ldots ,$ so that the connection
coefficients vanish at ${\bf x}.$ If $\delta $ is sufficiently
small, the difference between $(T_{AB,B} + T_{AN,N})$ and 
$(T_{AB;B} + T_{AN;N})$ is negligibly small, and it
follows that in the $\delta \to 0+$ limit 
\ba 
{\cal T}_{AB\vert B}+2T_{AN}(N=+\delta )=0,
\label{eqn:conservation:cal_T}
\ea 
where $\vert $ indicates the covariant derivative on 
$\Sigma ,$ or equivalently (in the $\delta \to 0+$ limit)
on the displaced surface $(\Sigma +\delta) $ just outside
the singular distribution, on which one imposes the Israel
matching condition.
Here $-T_{AN}$ represents the five-dimensional flux
of energy-momentum incident on the brane from the bulk 
(the direction $N$ is taken to be outward), or with a plus
sign it would represent energy-momentum flowing from the
brane into the bulk. 

We note that, alternatively, we could have derived 
eqn.~(\ref{eqn:conservation:cal_T}) directly from 
the Israel matching condition and the Gauss-Codazzi 
relations in eqn.~(\ref{eqn:Gauss:Codacci}).

Because we have assumed an empty bulk (except for the negative
cosmological constant), with 
$T_{ab} =-\Lambda^{(5)}g_{ab}$
outside of the brane, the flux contribution $2T_{AN}$ in
eqn.~(\ref{eqn:conservation:cal_T}) vanishes. This follows because 
the direction $A$ is normal to $N.$ 
Extra degrees of freedom (e.g., bulk scalars) in the
bulk would in general spoil the vanishing of $T_{AN}.$

In the discussion above, the (projected) brane stress-energy 
${\cal T}_{AB}$ has been shown to be a tensor field parallel to the
brane in both indices and conserved with respect to the
four-dimensional metric induced on the brane. For the components
parallel to the brane,
\ba
{\cal T}_{AB\vert B} ={\cal T}_{AB;B} =0,
\ea
where $\vert $ and $;$ denote covariant derivation with respect to the
four-dimensional metric induced on the brane and the full five-dimensional
metric, respectively.\footnote{ Note that, for example, with the
five-dimensional metric connection ${\cal T}_{AN;B}$ does not
necessarily vanish, even though ${\cal T}_{AN}$ is zero; however, in
the equations these terms are not included.}
The discussion so far in the present section applies equally well with and
without perturbations.

We now consider perturbations to linear order, splitting
${\cal T}_{AB} ={\cal T}^{(0)}_{AB} +{\cal T}^{(1)}_{AB}.$
At zeroth order, stress-energy conservation on the brane is expressed
as
${\cal T}^{(0)}_{AB\vert B} =0,$
where $\vert $ is the connection with respect to the unperturbed
metric induced on the brane. 

Stress-energy conservation to linear order is obtained 
by extracting the first-order terms
\ba
\left(g_{(0)}^{BC}-h^{BC}\right) \nabla ^{[g^{(0)}+h]}_B
\left( {\cal T}_{AC}^{(0)}+{\cal T}_{AC}^{(1)} \right) =0.
\ea
Here the superscript $[g^{(0)}+h]$ indicates that the 
covariant derivation is with respect to the perturbed
metric. Keeping only the first order terms of the above
equation gives
\ba
{\cal T}^{(1)}_{AB\vert B}
-h^{BC}~{\cal T}_{AC\vert B}^{(0)}
+\nabla ^{[h]}_B~{\cal T}_{AB}^{(0)}=0,
\ea
where $\vert $ denotes the connection of the unperturbed metric
induced on the brane.
Here the operator $\nabla ^{[h]}_A$ denotes the change in the 
metric connection due to the perturbation of the metric
from $g^{(0)}$ to $(g^{(0)}+h),$ accurate to linear order in $h.$
We may express this operator as
\ba 
\nabla ^{[h]}_A\hat e_B={\Gamma ^{[h]~C}}_{AB} ~\hat e_C ,
\ea
where the pseudo-Christoffel symbols above constitute a third-rank
tensor (and not a pseudo-tensor) with respect to the unperturbed
metric, and can be expressed as
\ba
{\Gamma ^{[h]}}_{CAB}={1\over 2}
\Bigl[ 
 h_{AC\vert B}
+h_{BC\vert A}
-h_{AB\vert C}
\Bigr] .
\label{eqn:pert_christoffel}
\ea
At a given point this relation can be demonstrated by choosing Fermi normal
coordinates there with respect to the unperturbed metric. Precisely at
this point  and in these coordinates, $g^{(0)}_{AB,C}=0$
and ${\Gamma ^{[g^{(0)}]~C}}_{AB}=0.$ It follows from the generally valid
expression
\ba
{\Gamma ^C}_{AB}
={1\over 2}g^{CD}\Bigl[ g_{AD,B}+g_{BD,A}-g_{AB,D}\Bigr] ,
\ea
that to linear order 
\ba 
{\Gamma ^{[h]~C}}_{AB}
={1\over 2}g_{(0)}^{CD}\Bigl[ h_{AD,B}+h_{BD,A}-h_{AB,D}\Bigr] .
\label{eqn:pert_christoffel:d}
\ea
Because the unperturbed Christoffel symbols all vanish, 
these ordinary derivatives
may be replaced by covariant derivatives, rendering equation 
eqn.~(\ref{eqn:pert_christoffel:d}) 
equivalent to eqn.~(\ref{eqn:pert_christoffel}). It follows that
\ba
\nabla ^{[h]}_B~{\cal T}^{(0)}_{AB}&=&
-{\Gamma ^{[h]~C}}_{BA}~{\cal T}^{(0)}_{CB}
-{\Gamma ^{[h]~C}}_{BB}~{\cal T}^{(0)}_{AC}\cr 
&=&
-{1\over 2}\Bigl(
 h_{BC\vert A}
+h_{CA\vert B}
-h_{BA\vert C}
\Bigr){\cal T}^{(0)}_{CB}\cr
&&-
{1\over 2}\Bigl(
 h_{BC\vert B}
+h_{CB\vert B}
-h_{BB\vert C}
\Bigr){\cal T}^{(0)}_{AC}\cr
&=&
-{1\over 2}\left[ 2{\cal T}^{(0)}_{AC}~h_{BC\vert B}
                  -{\cal T}^{(0)}_{AC}~h_{BB\vert C}
                  +{\cal T}^{(0)}_{CB}~h_{BC\vert A}
\right] .
\ea
The first-order terms of the stress-energy
conservation equation give 
\ba
&&\left[\nabla _{B} {\cal T}_{AB} \right]^{(1)}
=
{\cal T}^{(1)}_{AB\vert B} 
-h_{BC}~{\cal T}_{AC\vert B}^{(0)}
+\nabla ^{[h]}_B~{\cal T}_{AB}^{(0)} \cr
&=&
{\cal T}^{(1)}_{AB\vert B}
-h_{BC}~{\cal T}^{(0)}_{AC\vert B}
-          {\cal T}^{(0)}_{AC}~h_{BC\vert B}
-{1\over 2}{\cal T}^{(0)}_{BC}~h_{BC\vert A}
+{1\over 2}{\cal T}^{(0)}_{AC}~h_{BB\vert C} 
=0. \quad\quad
\label{eqn:conservation:T:O(h)}
\ea

\section{The Linearized Israel Matching Condition and Its Divergence}
\label{The Linearized Israel Matching Condition and Its Divergence}

In this section we show how to derive the form of the linearized
Israel matching condition for a $Z_2$ symmetric stress-energy
distribution of co-dimension one in a (4+1)-dimensional bulk spacetime.
The exact Israel matching condition
\ba
K_{AB}=-{\kappa ^2_{(5)}\over 2}
\left[ {\cal T}_{AB}-{1\over 3}g_{AB}{\cal T}_{CC}\right]
\label{eqn:Israel}
\ea
gives $K=+{\kappa ^2_{(5)}\over 6}{\cal T},$
which may be used to re-write eqn.~(\ref{eqn:Israel}) as 
\ba
K_{AB}-g_{AB}~K=-{\kappa _{(5)}^2\over 2}~{\cal T}_{AB},
\ea
giving us at first order the equation
\ba
K_{AB}^{(1)}
-g_{AB}^{(0)}\left(
K^{(1)}_{CC}
-K^{(0)}_{CD}~h_{CD}
\right)
-h_{AB}~K^{(0)}
=-{\kappa _{(5)}^2\over 2}~{\cal T}_{AB}^{(1)}.
\label{eqn:Israel:O(h)}
\ea

The first order perturbation in the extrinsic
curvature is [see Appendix \ref{appendix:ext:curv} for a derivation]
\ba
K^{(1)}_{AB}=
{1\over 2}\left[ h_{AB;N}-h_{AN;B}-h_{BN;A} \right]
+{1\over 2}K^{(0)}_{AB}~h_{NN}.
\ea
Using the supplementary boundary condition $h_{AN}=0,$
we may simplify the terms $h_{AN;B}+h_{BN;A}.$ Expanding
\ba
h_{AN;B}&=&\nabla _B \left[
 h_{CD}^{(c)}(\hat e_C\otimes \hat e_D)
+h_{NN}^{(c)}(\hat e_N\otimes \hat e_N)
\right] \cdot (\hat e_A\otimes \hat e_N)\cr
&=&h_{AD}^{(c)}~\hat e_N\cdot (\nabla _B \hat e_D)
  +h_{NN}^{(c)}~\hat e_A \cdot (\nabla _B \hat e_N)\cr
&=&
-K^{(0)}_{BD}~h_{AD}
+K_{AB}^{(0)}~h_{NN} ,
\label{eqn:h_{AN;B}}
\ea
we may now re-write
\ba
K^{(1)}_{AB}=
{1\over 2}\left[ h_{AB;N} +K_{BC}^{(0)}~h_{AC} +K_{AC}^{(0)}~h_{BC} 
\right]
-{1\over 2}K^{(0)}_{AB}~h_{NN},
\label{eqn:K:O(h)}
\ea
from which we find that
\ba
K^{(1)}_{CC}=
{1\over 2}h_{CC;N} +K_{CD}^{(0)}~h_{CD} -{1\over 2}K_{CC}^{(0)}~h_{NN}.
\ea

The next step is to re-express eqn.~{(\ref{eqn:K:O(h)}) in terms of
the trace-modified metric perturbation $\bar h_{ab}$ rather than
$h_{ab}.$ From 
\ba 
\bar h_{AB}
=h_{AB}-{1\over 2}g_{AB}^{(0)}\left( h_{CC}+h_{NN} \right) ,
\ea 
it follows that 
\ba 
\bar h_{AB;N}
=h_{AB;N}-{1\over 2}g_{AB}^{(0)}\left( h_{CC;N}+h_{NN;N} \right).
\ea 
We use the boundary condition [from eqn.~(\ref{eee})]
\ba
\bar h_{NN;N} +K^{(0)}_{AA}~\bar h_{NN} -K^{(0)}_{AB}~\bar h_{AB} =0
\ea
and 
\ba 
\bar h_{NN;N}
=h_{NN;N}-{1\over 2}g_{NN}^{(0)}\left( h_{CC;N}+h_{NN;N} \right) 
={1\over 2}\left( h_{NN;N}-h_{CC;N}\right) 
\ea
to obtain
\ba 
h_{NN;N}
=h_{CC;N}+2\bar h_{NN;N}
=h_{CC;N}-2K_{AA}^{(0)}~\bar h_{NN}+2K_{AB}^{(0)}~\bar h_{AB},
\ea
which gives 
\ba
\bar h_{AB;N}
=h_{AB;N}-g_{AB}^{(0)}~h_{CC;N}+g_{AB}^{(0)}
\left[K_{CC}^{(0)}~\bar h_{NN}-K_{CD}^{(0)}~\bar h_{CD}\right] .
\ea

Using eqns.~(\ref{eqn:h_{AN;B}}) and (\ref{eqn:K:O(h)}) and the relations
\ba
h_{ab}=\bar h_{ab}-{1\over 3}g^{(0)}_{ab}\bar h,\quad 
h=-{2\over 3}~\bar h,
\label{eqn:h-reverse}
\ea
we expand the left-hand side of the perturbed Israel matching
condition in eqn.~(\ref{eqn:Israel:O(h)}) as 
\ba
&&K_{AB}^{(1)}
- g_{AB}^{(0)}~ \left( K^{(1)}_{CC} -K^{(0)}_{CD}~h_{CD}\right)
-K^{(0)}~h_{AB}\cr
&=&{1\over 2}\left( h_{AB;N}+K^{(0)}_{AC}~h_{BC} +K^{(0)}_{BC}~h_{AC} \right)
-{1\over 2}K^{(0)}_{AB}~h_{NN}\cr
&&-g_{AB}^{(0)}~\left( {1\over 2}h_{CC;N}
+K^{(0)}_{CD}~h_{CD} -{1\over 2}K^{(0)}~h_{NN} \right)\cr
&&+g_{AB}^{(0)}~K^{(0)}_{CD}~h_{CD}-K^{(0)}~h_{AB}\cr
&=&
{1\over 2} \left( h_{AB;N}-g^{(0)}_{AB}~h_{CC;N}\right)
+{1\over 2} \left( K^{(0)}_{AC}~h_{BC} +K^{(0)}_{BC}~h_{AC}\right)\cr 
&&-{1\over 2}\left( K^{(0)}_{AB}-g^{(0)}_{AB}~K^{(0)}\right) h_{NN}
-K^{(0)}~h_{AB}\cr
&=&
{1\over 2}\left[
\bar h_{AB;N} + g^{(0)}_{AB}~\left( 
K^{(0)}_{CD}~\bar h_{CD}
-K^{(0)}~\bar h_{NN}
\right) \right]
\cr
&&+{1\over 2}K^{(0)}_{AC}~\left( 
\bar h_{BC} -{1\over 3}g^{(0)}_{BC}~\bar h\right) 
+{1\over 2}K^{(0)}_{BC}~\left( 
\bar h_{AC} -{1\over 3}g^{(0)}_{AC}~\bar h\right) \cr
&& -{1\over 2}\left[ K^{(0)}_{AB}-g^{(0)}_{AB}~K^{(0)}\right]
\left( \bar h_{NN}-{1\over 3}\bar h\right)\cr
&&-K^{(0)}\left( \bar h_{AB} -{1\over 3}g^{(0)}_{AB}~\bar h\right) \cr
&=&
{1\over 2}\bar h_{AB;N}
+ {1\over 2}K^{(0)}_{AC}~\bar h_{BC}
+ {1\over 2}K^{(0)}_{BC}~\bar h_{AC}
-K^{(0)}~\bar h_{AB}\cr
&&-{1\over 2}\left[ K^{(0)}_{AB}-g^{(0)}_{AB}~K^{(0)}\right]
\left( \bar h_{NN}-{1\over 3}\bar h\right)
-{1\over 3}K^{(0)}_{AB}~\bar h\cr
&&+g^{(0)}_{AB}
\left[
{1\over 2}K^{(0)}_{CD}~\bar h_{CD}
-{1\over 2}K^{(0)}~\bar h_{NN}
+{1\over 3}K^{(0)}~\bar h
\right] \cr
&=&
{1\over 2}\bar h_{AB;N}
+ {1\over 2}K^{(0)}_{AC}~\bar h_{BC}
+ {1\over 2}K^{(0)}_{BC}~\bar h_{AC}
-K^{(0)}~\bar h_{AB}\cr
&&
+{1\over 2}g^{(0)}_{AB}~K^{(0)}_{CD}~\bar h_{CD}
-{1\over 2}K^{(0)}_{AB}~\bar h_{NN}
-{1\over 6}K^{(0)}_{AB}~\bar h
+{1\over 6}g^{(0)}_{AB}~K^{(0)}\bar h.
\label{eqn:Israel:O(h):lhs}
\ea
We re-write eqn.~(\ref{eqn:Israel:O(h)}), obtaining for the linearized
Israel matching condition 
\ba
&&{1\over 2}\bar h_{AB;N}
+ {1\over 2}K^{(0)}_{AC}~\bar h_{BC}
+ {1\over 2}K^{(0)}_{BC}~\bar h_{AC}
-K^{(0)}~\bar h_{AB}\cr
&&
+{1\over 2}g^{(0)}_{AB}~K^{(0)}_{CD}~\bar h_{CD}
-{1\over 2}K^{(0)}_{AB}~\bar h_{NN}
-{1\over 6}K^{(0)}_{AB}~\bar h
+{1\over 6}g^{(0)}_{AB}~K^{(0)}\bar h 
=-{\kappa _{(5)}^2\over 2}~{\cal T}_{AB}^{(1)}. \quad\quad
\label{eqn:Israel:O(h):K}
\ea

We now proceed to take the divergence of eqn.~(\ref{eqn:Israel:O(h):K})
considered as a tensor field on the brane using the covariant
derivation $\tilde \nabla $ generated by the induced metric on the brane.
Noting that
\ba
\tilde \nabla_{B}\bar h_{AB;N}
&=&\bar h_{AB;N\vert B} \cr
&=&\bar h_{AB;NB} 
-K^{(0)}_{AB}~\bar h_{NB;N}
-K^{(0)}_{BB}~\bar h_{AN;N}
+K^{(0)}_{CB}~\bar h_{AB;C}, 
\ea
we find the divergence of the left-hand side of eqn.~(\ref{eqn:Israel:O(h):K})
is
\ba
&&
\nabla_{B}\left[ 
K_{AB}^{(1)}
- g_{AB}^{(0)}~ \left( K^{(1)}_{CC} -K^{(0)}_{CD}~h_{CD}\right)
-K^{(0)}~h_{AB} 
\right] \cr
&=&
{1\over 2}\bar h_{AB;NB} \cr
&&
-{1\over 2}K^{(0)}_{AB}~\bar h_{NB;N}
-{1\over 2}K^{(0)}_{BB}~\bar h_{AN;N}
+{1\over 2}K^{(0)}_{CB}~\bar h_{AB;C} \cr
&&
+{1\over 2}K^{(0)}_{BC}~\bar h_{AC;B}
+{1\over 2}K^{(0)}_{AC}~\bar h_{BC;B}
+{1\over 2}K^{(0)}_{BC}~\bar h_{BC;A}
-          K^{(0)}_{}~\bar h_{AB;B}
-{1\over 2}K^{(0)}_{AB}~\bar h_{NN;B}  \cr
&&
-{1\over 6}K^{(0)}_{AB}~\bar h_{;B}
+{1\over 6}K^{(0)}_{}~\bar h_{;A}\cr
&&
+{1\over 2}K^{(0)}_{BC;B}~\bar h_{AC}
+{1\over 2}K^{(0)}_{AC;B}~\bar h_{BC}
+{1\over 2}K^{(0)}_{BC;A}~\bar h_{BC} 
-          K^{(0)}_{;B}~\bar h_{AB}
-{1\over 2}K^{(0)}_{AB;B}~\bar h_{NN}\cr
&&
-{1\over 6}K^{(0)}_{AB;B}~\bar h
+{1\over 6}K^{(0)}_{;A}~\bar h~.
\label{eqn:div:Israel:O(h):lhs}
\ea
We next compute the divergence of the right-hand side of 
eqn.~(\ref{eqn:Israel:O(h):K}),
using the result from the stress-energy conservation at first 
subleading order 
[eqn.~(\ref{eqn:conservation:T:O(h)})] 
\ba
{\cal T}^{(1)}_{AB\vert B} 
&=&
{\cal T}_{AC\vert B}^{(0)}~h_{BC} 
-\nabla ^{[h]}_B {\cal T}_{AC}^{(0)} \cr
&=&
{\cal T}^{(0)}_{AC\vert B}~h_{BC}
+          {\cal T}^{(0)}_{AC}~h_{BC\vert B}
+{1\over 2}{\cal T}^{(0)}_{BC}~h_{BC\vert A}
-{1\over 2}{\cal T}^{(0)}_{AC}~h_{BB\vert C}.
\label{oneohsix}
\ea
Because of the boundary condition $h_{AN}=0,$ $h_{AB\vert C}=h_{AB;C}.$
The zeroth-order Israel matching condition
\ba
K_{AB}^{(0)}-g_{AB}^{(0)}~K^{(0)}
=-{\kappa _{(5)}^2\over 2}~{\cal T}_{AB}^{(0)}
\label{eqn:Israel:O(1)}
\ea
allows one to express eqn.~(\ref{oneohsix}) as  
\ba
-{\kappa _{(5)}^2 \over 2}{\cal T}^{(1)}_{AB\vert B} 
&=&
\left[ K_{AC;B}^{(0)}-g_{AC}^{(0)}~K^{(0)}_{;B} \right]h_{BC} 
\cr &&
+K_{AC}^{(0)}~h_{BC; B}
-K^{(0)}~h_{BA; B}
+{1\over 2}K_{BC}^{(0)}~h_{BC; A}
-{1\over 2}K_{AC}^{(0)}~h_{BB; C}.
\ea
Using the definition of the trace-modified metric perturbation
[eqn.~(\ref{eqn:h-reverse})], we find that
\ba
-{\kappa _{(5)}^2 \over 2}{\cal T}^{(1)}_{AB\vert B} 
&=&
\left[ K_{AC; B}^{(0)}-g_{AC}^{(0)}~K_{; B}^{(0)} \right]
 \left(\bar h_{BC}-{1\over 3}g_{BC}^{(0)}~\bar h \right) \cr
&&+
K_{AC}^{(0)}~\bar h_{BC;B}
-K^{(0)}~\bar h_{BA;B}
+{1\over 2}K_{BC}^{(0)}~\bar h_{BC;A}
-{1\over 2}K_{AC}^{(0)}~\bar h_{BB;C} \cr
&&
+{1\over 3}K_{AC}^{(0)}~\bar h_{;C}
+{1\over 6}K^{(0)}~\bar h_{;A},
\label{eqn:div:T:O(h)}
\ea
and noting that $\bar h =\bar h_{BB} +\bar h_{NN}$
we can re-write eqn.(\ref{eqn:div:T:O(h)}) as
\ba
-{\kappa _{(5)}^2 \over 2}{\cal T}^{(1)}_{AB\vert B} 
&=&
\left[ K_{AC; B}^{(0)}-g_{AC}^{(0)}~K_{; B}^{(0)} \right]
 \left(\bar h_{BC}-{1\over 3}g_{BC}^{(0)}~\bar h \right) \cr
&&
+K_{AC}^{(0)}~\bar h_{BC;B}
-K^{(0)}~\bar h_{BA;B}
+{1\over 2}K_{BC}^{(0)}~\bar h_{BC;A}
+{1\over 2}K_{AC}^{(0)}~\bar h_{NN;C} \cr
&&
-{1\over 6}K_{AC}^{(0)}~\bar h_{;C}
+{1\over 6}K^{(0)}~\bar h_{;A} .
\label{eqn:div:Israel:O(h):rhs}
\ea
Subtracting eqn.~(\ref{eqn:div:Israel:O(h):rhs}) from
eqn.~(\ref{eqn:div:Israel:O(h):lhs})
we find that vanishing of the divergence of the Israel matching condition and
stress-energy conservation on the brane imply
\ba
&&
+{1\over 2}\bar h_{AB;NB} \cr
&&
-{1\over 2}K^{(0)}_{AB}~\bar h_{NB;N}
-{1\over 2}K^{(0)}_{BB}~\bar h_{AN;N}
-{1\over 2}K_{AC}^{(0)}~\bar h_{BC;B}
\cr
&&
+K_{BC}^{(0)}~\bar h_{AC;B}
-K_{AB}^{(0)} ~\bar h_{NN;B} \cr
&&
+{1\over 2}K_{BC;B}^{(0)}~\bar h_{AC}
-{1\over 2}K_{AC;B}^{(0)}~\bar h_{BC}
+{1\over 2}K_{BC;A}^{(0)}~\bar h_{BC}
-{1\over 2}K_{AB;B}^{(0)} ~\bar h_{NN} \cr
&&
+{1\over 6}K_{AB;B}^{(0)}~\bar h
-{1\over 6}K_{;A}^{(0)}~\bar h =0.
\label{eqn:div:Israel:O(h)}
\ea
The zeroth order Israel matching condition
[eqn.~(\ref{eqn:Israel:O(1)})]
and stress-energy conservation at zeroth order 
${\cal T}^{(0)}_{AB\vert B} =0$
yield
\ba
K^{(0)}_{AB;B} -K^{(0)}_{;A} =0,
\label{eqn:div:Israel:O(1)}
\ea
which may be used to rewrite eqn.~(\ref{eqn:div:Israel:O(h)}) 
as follows
\ba
&&
{1\over 2}\bar h_{AB;NB} \cr
&&
-{1\over 2}K^{(0)}_{AB}~\bar h_{NB;N}
-{1\over 2}K^{(0)}_{BB}~\bar h_{AN;N}
-{1\over 2}K_{AC}^{(0)}~\bar h_{BC;B}
\cr
&&
+K_{BC}^{(0)}~\bar h_{AC;B}
-K_{AB}^{(0)} ~\bar h_{NN;B} \cr
&&
+{1\over 2}K_{BC;B}^{(0)}~\bar h_{AC}
-{1\over 2}K_{AC;B}^{(0)}~\bar h_{BC}
+{1\over 2}K_{BC;A}^{(0)}~\bar h_{BC}
-{1\over 2}K_{AB;B}^{(0)} ~\bar h_{NN} 
 =0.
\label{onethirteen}
\ea
We note that eqn.~(\ref{onethirteen}) resulted
from assuming the Israel matching condition
and stress-energy conservation on the brane. 
Our object was to show that $V_{A;N},$ given
in eqn.~(\ref{eqn:va;n}), vanishes. 
We note that eqn.~(\ref{onethirteen}) equals
precisely one half the right-hand side of
(\ref{eqn:va;n}). Hence it follows that
$V_{A;N}=0.$

\section{Discussion}
\label{Discussion}

We summarize our main results as follows. 
Throughout the paper we have assumed Lorentz gauge in the bulk 
for the linearized metric perturbations---that is
\be 
\bar h_{ab;b}=0,
\label{lgc2}
\ee
where
in terms of the linear metric perturbation
$h_{ab}$
\be 
\bar h_{ab}=h_{ab}
  -{1\over 2}g_{ab}^{(0)}~\Bigl( g^{(0)~cd}~h_{cd}\Bigr) .
\ee 
In the bulk the Lorentz gauge condition is consistent as long
as the bulk stress-energy is conserved. 
We have demonstrated that
the auxiliary boundary conditions on the brane
\be
\bar h_{AN}=0
\label{zza}
\ee 
and
\be
\bar h_{NN,N}+K_{AA}~\bar h_{NN}-K_{AB}~\bar h_{AB}=0
\label{zzb}
\ee 
together with the Israel matching condition 
sourced by a conserved stress-energy on the brane ${\cal T}_{AB}$
ensure that the waves reflecting from the boundary
or emanating from sources there respect the
bulk Lorentz gauge condition. 

The bulk wave equation in $(4+1)$ dimensions 
[given in eqn. (\ref{eqn:h:wave:source}), which simplifies to
eqn. (\ref{eqn:h:wave:AdS5}) for an $AdS^5$ bulk] 
propagates 15 components. On the boundary
the Israel matching condition provides
10 of the 15 required boundary conditions.
Eqns.~(\ref{zza}) and (\ref{zzb}) provide the remaining
$4+1$ 
boundary conditions, respectively,
required to render reflection off the boundary
unique and consistent.

The Lorentz gauge condition in the bulk [eqn.~(\ref{lgc2})]
does not fix the gauge on the brane. Rather, once a choice
of gauge on the brane has been chosen, it provides 
a prescription 
for continuing this gauge choice into the bulk. 
In (4+1) dimensions the Lorentz gauge condition
allows exactly five longitudinal (pure gauge)
modes in the bulk. Four of these correspond
to reparameterization of the brane. The one
remaining mode corresponds to normal displacements
of the brane, which become physical if the 
brane is fixed to its unperturbed position
with respect to the unperturbed background 
spacetime.\\

{\bf Acknowledgements:} 
MB thanks Mr Denis Avery and the CNRS for support. CC thanks
the Gulbenkian Foundation for support and the Laboratoire de
Physique Theorique at Orsay for its hospitality. We thank John
Stewart for help using REDUCE and Christos Charmousis 
and especially Pierre Bin\'etruy
for useful discussions.

\appendix

\section{Bulk Metric Perturbation Evolution Equations in Lorentz Gauge}
\label{appendix:evo:eqn:h}

In this appendix we derive the evolution equation for the metric perturbation
$\bar h_{ab}$ and for the violation of the gauge. We assume the Einstein
equation 
$G_{ab} =\left[ R_{ab}-{1\over 2}g_{ab}R\right] =(8\pi G)T_{ab}$
and the Lorentz gauge condition $\bar h_{ab;b} =0$.
We also demonstrate the consistency of the Lorentz gauge condition
in the bulk assuming that the bulk stress-energy is conserved. 
For the linearized perturbations
\ba
G^{(1)}_{ab} =R^{(1)}_{ab} 
-{1\over 2}\left[ g^{(0)}_{ab} R^{(1)} +h_{ab} R^{(0)} \right]
=(8\pi G)T^{(1)}_{ab},
\label{eqn:Einstein:O(h)}
\ea
where $R^{(1)} = g^{(0)~cd}~R^{(1)}_{cd} -h^{cd}~R^{(0)}_{cd}.$
The Riemann tensor is given by
\ba
{R^{a}}_{bcd} =
 {\Gamma^{a}}_{bd,c} 
-{\Gamma^{a}}_{bc,d} 
+{\Gamma^{a}}_{ec}~{\Gamma^{e}}_{bd}
-{\Gamma^{a}}_{ed}~{\Gamma^{e}}_{bc}
\ea
where
${\Gamma^{a}}_{bc} ={1\over 2}g^{ad} 
 \left[ g_{dc,b} +g_{bd,c} -g_{bc,d} \right] $
can be decomposed as 
$\Gamma^{a}_{bc} = {\Gamma^{[g^{(0)}]}} ^{a}_{bc} +{\Gamma^{[h]}} ^{a}_{bc},$
where
${\Gamma^{[h]}} ^{a}_{bc} ={1\over 2}g_{(0)}^{ad} 
 \left[ h_{dc;b} +h_{bd;c} -h_{bc;d} \right]. $
The linear perturbation of the Riemann 
tensor is
${R^{(1)}}^{a}_{bcd} =
{\Gamma^{[h]}}^{a}_{bd;c} -{\Gamma^{[h]}}^{a}_{bc;d},$
and for the Ricci tensor
\ba
R^{(1)}_{ab} \equiv 
{R^{(1)}}^{c}_{acb}
={1\over 2} \left[ -h_{;ab} -h_{ab;cc} +h_{ca;bc} +h_{cb;ac} \right].
\ea
From the definition of $\bar h_{ab}$ [eqn.~(\ref{eqn:hbar})], 
$h_{ab} =\bar h_{ab} -(d-2)^{-1}g^{(0)}_{ab}\bar h,$
$h =-2(d-1)^{-1}\bar h .$
Expressing eqn.~(\ref{eqn:Einstein:O(h)}) in terms of $\bar h_{ab},$ we
find that 
\ba
G^{(1)}_{ab}
&=& 
{1\over 2}\left[ 
-\bar h_{ab;cc} +\bar h_{ca;bc} +\bar h_{cb;ac} 
-g^{(0)}_{ab}\bar h_{cd;cd} \right] 
-{1\over 2}\left[
-g^{(0)}_{ab}\bar h_{cd}+\bar h_{ab}g^{(0)}_{cd} \right]R^{(0)}_{cd} 
\cr &=&
(8\pi G)~T^{(1)}_{ab}.
\label{eqn:G:O(h)}
\ea
Using
$\bar h_{ab;cd} -\bar h_{ab;dc}
=R_{aedc}^{(0)}\bar h_{eb} +R_{bedc}^{(0)}\bar h_{ae},$
and imposing the Lorentz gauge condition gives
\ba
\Box \bar h_{ab}
- \left[
g^{(0)}_{bc} \bar h_{da}
+g^{(0)}_{ac} \bar h_{db}
+g^{(0)}_{ab} \bar h_{cd}
-g^{(0)}_{cd} \bar h_{ab}
\right] R^{(0)}_{cd}
+2R^{(0)}_{acbd}\bar h_{cd}
=-(16\pi G)~T^{(1)}_{ab}.
\label{eqn:h:wave}
\ea
as the wave equation in Lorentz gauge.

We next demonstrate the consistency of the Lorentz gauge condition 
with the above evolution equation. To this end, we define the field
\be
V_a=\bar h_{ab;b}
\ee
quantifying any possible violation of the Lorentz gauge
condition and show that $V_a$ satisfies a homogeneous
wave equation---that is, one with no non-vanishing sources,
as long as the bulk stress-energy is conserved 
(i.e., $T_{ab;b}=0$). The fact that the wave equation
for $V_a$ is homogeneous implies that if $V_a$
and its normal time derivative on an initial past
Cauchy surface vanishes, then $V_a$ vanishes
everywhere in the future.

We now take the divergence of eqn.~(\ref{eqn:h:wave}).
The identities
\ba
D_{ab;mn} -D_{ab;nm} 
&=&R_{acnm}{D^{c}}_{b} +R_{bcnm}{D_{a}}^{c},\cr
S_{abc;mn} -S_{abc;nm}
&=&R_{adnm}{S^{d}}_{bc} +R_{bdnm}{{S_{a}}^{d}}_{c} +R_{cdnm}{S_{ab}}^{d}
\ea
give
\ba
(\Box \bar h_{ab})_{;b}
&=& \bar h_{ab;ccb} 
= \bar h_{ab;cbc} 
+R_{adbc}\bar h_{db;b} +R_{bdbc}\bar h_{ad;c} +R_{cdbc}\bar h_{ab;d}
\cr
&=& \bar h_{ab;bcc}
+( R_{adbc}\bar h_{db} +R_{bdbc}\bar h_{ad} )_{;c}
+R_{adbc}\bar h_{db;b} +R_{bdbc}\bar h_{ad;c} +R_{cdbc}\bar h_{ab;d}
\cr
&=& \Box \bar h_{ab;b}
-R_{adcb;c}\bar h_{db} +R_{dc;c}\bar h_{ad}
-2R_{adcb}\bar h_{db;c} +2R_{dc}\bar h_{ad;c} -R_{db}\bar h_{ab;d},
\quad\quad\quad
\label{eqn:box:h}
\ea 
which can be used to take the divergence of eqn.~(\ref{eqn:h:wave})
\ba
(\Box \bar h_{ab} )_{;b} 
&-&\left[ 
g^{(0)}_{bc}\bar h_{da;b}  
+g^{(0)}_{ab}\bar h_{cd;b} 
\right]R^{(0)}_{cd} 
+2R^{(0)}_{adbc}\bar h_{dc;b} \cr
&-&\left[ 
g^{(0)}_{bc}\bar h_{da} +g^{(0)}_{ac}\bar h_{db} 
+g^{(0)}_{ab}\bar h_{cd} -g^{(0)}_{cd}\bar h_{ab}
\right]R^{(0)}_{cd;b} 
+2R^{(0)}_{adbc;b}\bar h_{dc} 
= -(16\pi G)T^{(1)}_{ab;b}.\quad \quad
\ea
Using eqn.~(\ref{eqn:box:h}), we obtain
\ba
&&\Box \bar h_{ab;b}
-\bar h_{cd;a}R^{(0)}_{cd} 
-\left[ 
g^{(0)}_{ac}\bar h_{db} +g^{(0)}_{ab}\bar h_{cd} -g^{(0)}_{cd}\bar h_{ab} 
\right]R^{(0)}_{cd;b} 
+R^{(0)}_{adbc;b}\bar h_{dc} 
= -(16\pi G)T^{(1)}_{ab;b}. \quad \quad \quad
\label{eqn:div:h:wave:R}
\ea
The zeroth order Ricci tensor can be replaced using
\ba
R^{(0)}_{ab} =(8\pi G) \left[
T^{(0)}_{ab} -{1\over{d-1}}g^{(0)}_{ab}T^{(0)} 
\right].
\label{eqn:Ricci:O(1)}
\ea
The Bianchi indentity
$R^{(0)}_{ac[bd;b]} =R^{(0)}_{bd[ac;b]}
=R^{(0)}_{bdac;b} +R^{(0)}_{bdcb;a} +R^{(0)}_{bdba;c} =0$
gives the relation
$ R^{(0)}_{bdac;b} -R^{(0)}_{dc;a} +R^{(0)}_{da;c} =0,$ 
which can be used to re-write
\ba
&&R^{(0)}_{acbd;b}\bar h_{cd} =R^{(0)}_{bdac;b}h_{cd}
=\left( R^{(0)}_{dc;a} -R^{(0)}_{da;c} \right)\bar h_{cd}\cr 
&&\quad =(8\pi G)\left[ 
{\cal T}^{(0)}_{dc;a} -{\cal T}^{(0)}_{da;c}
-{1\over {d-1}}\left( 
g^{(0)}_{dc}{\cal T}^{(0)}_{;a}
-g^{(0)}_{da}{\cal T}^{(0)}_{;c} 
\right)
\right]\bar h_{cd},
\label{eqn:Bianchi}
\ea
giving
\ba
\Box \bar h_{ab;b}
=-(16\pi G)~\left[
{\cal T}^{(1)}_{ab;b}-{1\over 2}{\cal T}^{(0)}_{cd}\bar h_{cd;a}
+{1\over 2(d-1)}{\cal T}^{(0)}\bar h_{;a}
-{\cal T}^{(0)}_{ad;b}\bar h_{db}
\right] .
\label{eqn:div:h:wave:T}
\ea
Using the first-order stress-energy
conservation equation previously obtained in 
eqn.~(\ref{eqn:conservation:T:O(h)}) 
\ba
\left[\nabla _{b} {\cal T}_{ab} \right]^{(1)}
&=&
{\cal T}^{(1)}_{ab;b}
-          {\cal T}^{(0)}_{ac;b}h_{bc}
-          {\cal T}^{(0)}_{ac}~h_{bc;b}
-{1\over 2}{\cal T}^{(0)}_{bc}~h_{bc;a}
+{1\over 2}{\cal T}^{(0)}_{ac}~h_{bb;c} \cr
&=&
{\cal T}^{(1)}_{ab;b}
-          {\cal T}^{(0)}_{ac;b}\bar h_{bc}
-{1\over 2}{\cal T}^{(0)}_{bc}\bar h_{bc;a}
+{1\over 2}{1\over {d-1}}{\cal T}^{(0)}\bar h_{;a}
=0
\ea
we obtain 
$\Box V_a=
\Box \bar h_{ab;b} =0,$
proving the consistency of Lorentz gauge at linear order
provided that stress-energy is conserved.

\section{Derivation of the Perturbed Israel Matching Condition}
\label{appendix:imc}

We derive the Israel matching conditions for a
surface of codimension one having singular distribution of
stress-energy embedded in a $(d+1)$-dimensional bulk spacetime.
The Gauss-Codacci relations \cite{capovilla}, \cite{spivak}
\ba
R_{ABCD}
&=&R^{(ind)}_{ABCD} +K_{AD}K_{BC} -K_{AC}K_{BD}, \cr
R_{NBCD}
&=&K_{BC; D} -K_{BD; C}, \cr
R_{NBND}
&=&K_{BC}K_{DC} -K_{BD,N},
\label{eqn:Gauss:Codacci}
\ea
decompose
the Riemann tensor into a components tangent and normal to the brane.
The Ricci tensor and scalar are given by
\ba
R_{AB} 
&=&R_{CACB} +R_{NANB} 
=R^{(ind)}_{AB} +2K_{AC}K_{CB} -K_{AB}K -K_{AB,N}, \cr
R_{NA}
&=& K_{BA; B} -K_{; A}, \cr
R_{NN}
&=&R_{NCNC} =K_{AB}K_{AB} -K_{,N},
\ea
and 
\ba
R
&=&R_{AB}g^{AB} +R_{NN}g^{NN} 
=R^{(ind)} +3K_{AB}K_{BA} -K^2 -2K_{,N}. 
\ea
The Einstein equations decompose similarly
\ba
G_{AB} 
&=& G^{(ind)}_{AB} +2K_{AC}K_{CB} -K_{AB}K -K_{AB,N} \cr
&&-{1\over 2}g^{(ind)}_{AB}\left[ 3K_{CD}K_{DC} -K^2 -2K_{,N} \right], \cr
G_{AN}
&=& K_{AB; B} -K_{; A}, \cr
G_{NN}
&=&
{1\over 2}
 \left[-R^{(ind)} -K_{AB}K_{BA} +K^2\right],
\ea
where $K=g_{(ind)}^{AB}K_{AB}.$ 

At the brane itself, where the metric is continuous but its
normal derivative may suffer a jump, the singular contribution to
$G_{AB}$ 
arises from the terms
\ba
-K_{AB,N} + g^{(ind)}_{AB}K_{,N}, 
\ea
which when integrated across the brane give
\ba
\left[K_{AB}-Kg^{(ind)}_{AB}\right|^{+\delta}_{-\delta}=
-\kappa_{(d+1)}^2 {\cal T}_{AB},
\label{eqn:israel}
\ea
where we define
four-dimensional projected
stress-energy tensor  
\ba
{\cal T}_{AB} =\int^{+\delta}_{-\delta}dN~T_{AB},
\ea
where $\kappa_{(d+1)}^2=8\pi {G}_{(d+1)}=M_{(d+1)}^{-(d-2)}.$
We can trace-modify eqn.~(\ref{eqn:israel}) to obtain
\ba
\left[K_{AB} \right|^{+\delta}_{-\delta}=
-\kappa_{(d+1)}^2\left[ {\cal T}_{AB}-{1\over d-1}{\cal T}g^{(ind)}_{AB} 
\right].
\ea

\section{Derivation of Extrinsic Curvature Perturbation}
\label{appendix:ext:curv}

To compute the perturbation in the extrinsic curvature
derived from a perturbation in the metric
we use the relation 
\ba
K_{AB}={+1\over 2}\left[ 
{\cal L}_{\left( \overline{N}+\delta \overline{N}\right) }.
\left(
\underline{\underline{g}}^{(0)}
+\underline{\underline{h}}
\right) 
\right] _{AB}.
\ea
This formulation in terms of the Lie derivative circumvents having to
consider how covariant differentiation is modified by the metric
perturbation. 
It follows that
\ba
K^{(1)}_{AB}=
{1\over 2}\left[
{\cal L}_{\overline{N}}~ \underline{\underline{h}}
+{\cal L}_{\delta \overline{N}}~ \underline{\underline{g}}^{(0)}
\right] _{AB}.
\ea
For any doubly covariant tensor field
$\underline{\underline{\omega}}$ and contravariant vector field
$\overline{U},$ the following relation 
\ba
\left( {\cal L}_{\overline{U}}~\underline{\underline{\omega}} \right)_{ab}
= U^c~\omega_{ab;c} 
+ \omega_{ac}~{U^c}_{;b}
+ \omega_{cb}~{U^c}_{;a}
\ea
holds regardless of the choice of covariant derivation
$;$ as long as the connection is torsion free.
This property reflects the fact that the Lie 
derivative is defined solely in terms of the flow induced by
$\overline{U}$, and consequently is independent of
any metric or affine, torsion free structure defined
on the manifold. We find it
most convenient to set $;$ to the metric 
connection defined by the unperturbed metric
$g^{(0)}_{ab},$ in accord with the convention of the rest
of this paper. To zeroth order one has $K^{(0)}_{AB}=N_{A;B}.$

We first calculate $\delta \overline{N}$ by requiring that
\ba 
\Bigl( N^a +\delta N^a \Bigr) 
\Bigl( g^{(0)}_{ab} +h_{ab} \Bigr)
\Bigl( N^b +\delta N^b \Bigr) 
=+1,\quad 
\Bigl( N^a+\delta N^a \Bigr)
\Bigl( g^{(0)}_{ab}+h_{ab} \Bigr) X_A^b=0,
\ea 
for all directions $X_A^b$ parallel to the brane.
Here  $X_A^b$ is a contravariant unit vector so that
$g_{(0)}^{ab} =N^{a}N^{b} +\sum_{A}X_A^a X_A^b.$
It follows that
$\delta N^N ={-1\over 2}h_{NN}, 
\delta N^A =-{h_{N}}^A,$
which may be re-written as 
\ba
\delta N^a
={1\over 2}\Bigl( N^c h_{cd}N^d \Bigr)N^a
-g_{(0)}^{ab}h_{bc}N^c,
\ea
so that
\ba
\delta N_{a;b}
&=&
{1\over 2}h_{NN}N_{a;b}
+{1\over 2}N^{c}h_{cd;b}N^{d}N_{a}
+N^{c}h_{cd}{N^{d}}_{;b}N_{a}
-h_{ac;b}N^c
-h_{ac}{N^{c}}_{;b} \cr
&=&
{1\over 2}h_{NN}K^{(0)}_{ab}
+{1\over 2}h_{NN;b}N_{a}
+h_{CN}K^{(0)}_{Cb}N_{a}
-h_{aN;b}
-h_{aC}K^{(0)}_{Cb} .
\ea
We now obtain the first order extrinsic
curvature perturbation 
\ba
K_{AB}^{(1)}&=&{+1\over 2}\left[
 {\cal L}_{\overline{N}}~
         \underline{\underline{h}}
+{\cal L}_{\delta \overline{N} }~
         \underline{\underline{g}}^{(0)}
\right] _{AB}\cr
&=&{1\over 2}X^a_A X^b_B N^c h_{ab;c}
+{1\over 2}X^a_A X^b_B \left[
h_{ac}{N^{c}}_{;b}+h_{bc}{N^{c}}_{;a}\right] \cr
&&+{1\over 2}X^a_A X^b_B \left[
(\delta N)_{a;b}+(\delta N)_{b;a})
\right] \cr 
&=&
 {1\over 2}h_{AB;N}
+{1\over 2}K^{(0)}_{BC}~h_{CA}
+{1\over 2}K^{(0)}_{AC}~h_{CB}\cr
&&+{1\over 2}h_{NN}~K^{(0)}_{AB}
-{1\over 2}h_{AN;B}-{1\over 2}h_{BN;A}
-{1\over 2}K^{(0)}_{BC}~h_{CA}
-{1\over 2}K^{(0)}_{AC}~h_{CB}\cr
&=& {1\over 2}h_{AB;N}-{1\over 2}h_{AN;B}-{1\over 2}h_{BN;A}
+{1\over 2}h_{NN}~K^{(0)}_{AB} .
\ea

\end{document}